\documentclass[a4paper]{article}

\usepackage[left = 3 cm, right = 3 cm]{geometry}
\usepackage[T1]{fontenc} 
\usepackage[utf8]{inputenc}
\usepackage{lmodern}
\usepackage{bm} 
\usepackage{amsmath, amsfonts, amssymb}
\usepackage{graphicx}
\usepackage{hyperref}
\usepackage{upgreek}
\usepackage[super, sort&compress]{natbib}
\usepackage{authblk}
\usepackage{todonotes, soul, xcolor}
\usepackage{lineno}
\usepackage{setspace}
\usepackage{multirow}
\usepackage{makecell}
\usepackage{rotating}

\author[1, *]{Zolt\'{a}n Tajkov}
\author[2]{D\'{a}niel Nagy}
\author[1]{Konr\'{a}d Kandrai}
\author[4]{J\'{a}nos Koltai}
\author[2, 3]{L\'{a}szl\'{o} Oroszl\'{a}ny}
\author[1]{P\'{e}ter S\"{u}le}
\author[1]{Zsolt E. Horv\'{a}th}
\author[1]{P\'{e}ter Vancs\'{o}}
\author[1]{Levente Tapaszt\'{o}}
\author[1, *]{P\'{e}ter Nemes-Incze}

\affil[1]{Centre for Energy Research, Institute of Technical Physics and Materials Science, 1121 Budapest, Hungary}
\affil[2]{Department of Physics of Complex Systems, ELTE Eötvös Loránd University, 1117 Budapest, Hungary}
\affil[3]{Budapest University of Technology and Economics, 1111 Budapest, Hungary}
\affil[4]{ELTE E\"{o}tv\"{o}s Lor\'{a}nd University, Department of Biological Physics, 1117 Budapest, Hungary}
\affil[*]{\small \emph{corresponding author, email: tajkov.zoltan@ek-cer.hu, nemes.incze.peter@ek-cer.hu}}

\title{Revealing the topological phase diagram of ZrTe$_5$ using the complex strain fields of microbubbles}

\begin{document}

\maketitle

\doublespacing

\textbf{
Topological materials host robust properties, unaffected by microscopic perturbations, owing to the global topological properties of the bulk electron system.
Materials in which the topological invariant can be changed by easily tuning external parameters are especially sought after.
Zirconium pentatelluride (ZrTe$_5$) is one of a few experimentally available materials that reside close to the boundary of a topological phase transition, allowing the switching of its invariant by mechanical strain.
Here, we unambiguously identify a topological insulator - metal transition as a function of strain, by a combination of \emph{ab initio} calculations and direct measurements of the local charge density.
Our model quantitatively describes the response to complex strain patterns found in bubbles of few layer ZrTe$_5$ without fitting parameters, reproducing the mechanical deformation dependent closing of the band gap observed using scanning tunneling microscopy.
We calculate the topological phase diagram of ZrTe$_5$ and identify the phase at equilibrium, enabling the design of device architectures which exploit the unique topological switching characteristics of the system.
}


\section*{Introduction}

The paradigm of topological phases has permeated much of contemporary condensed matter physics \cite{kosterlitz1973ordering,laughlin1981quantized,hasan2010colloquium,armitage2018weyl}. This fundamentally new way of classifying systems according to global topological properties rather than a local order parameter yielded a deeper understanding of a host of peculiarly robust phenomena \cite{asboth2016short}.
At the heart of these phenomena lies the bulk-boundary correspondence, which guarantees robust states localized at the perimeter of the topological materials.
These boundary states, in turn, might be used as tools for measuring fundamental constants\cite{maciejko2010topological}, as components in thermoelectrics\cite{xu2017topological} or in spintronics devices\cite{brune2012spin}. 

Time reversal symmetric band insulators can be characterized by a $\mathbb{Z}_2$ index and can be classified into three phases in three dimensions.
A normal, or trivial insulating phase, a weak topological insulating (WTI) and strong topological insulating (STI) phase~\cite{wieder2021topological}.
These phases are separated by a metallic Dirac or Weyl semimetal phase\cite{bouhon2020non}. In order to change the $\mathbb{Z}_2$ it is necessary to close and reopen the bandgap through one of these metallic phases\cite{asboth2016short}.
The transition-metal pentatelluride ZrTe$_5$ is an excellent material to investigate topological phase transitions because it lies close to the boundary between a STI and WTI~\cite{mutch2019evidence}.
Additionally, the material has been widely studied due to its numerous exotic properties.
In the monolayer limit it is predicted to be large bandgap quantum spin Hall insulator~\cite{weng2014transition}, it has high mobility Dirac carriers~\cite{wang2020facile,liu2016zeeman}, it shows the chiral magnetic effect\cite{xie2021electron,li2016chiral} and the 3D quantum Hall effect~\cite{tang2019three} as well as multiple superconducting phases have been discovered under high pressure~\cite{zhou2016pressure}.

The topological nature of the bulk ZrTe$_5$ has not been unambiguously identified, with some experiments pointing to a STI and others to a WTI or a semimetal phase, as reviewed by Monserrat et al~\cite{monserrat2019unraveling}.
For example, angle resolved photoemission studies show evidence of a STI phase~\cite{manzoni2016evidence}, as well as the WTI case \cite{xiong2017three}.
These experiments are supported by first-principles calculations that also show the very same pattern of contrasting results.

In our contribution, we start from \textit{ab initio} calculations of ZrTe$_5$ and calculate the complete elastic response of the crystal by obtaining the elastic tensor elements, using a new approach compared to the literature\cite{de2015charting}.
We validate our \textit{ab initio} method by reproducing the closing of the band gap at the perimeter of bubbles formed by few layer ZrTe$_5$ on Au(111).
Measuring the surface charge density of bubbles provides an almost ideal experimental platform to validate our calculations, because bubbles of few layer van der Waals materials provide a varied deformation landscape~\cite{khestanova2016universal,petHo2019moderate,Lu2012-tg,Levy2010-bi}.
This deformation landscape leads to a local perturbation of the charge density that can be directly mapped, using STM.
Because STM directly measures the surface charge density, it does not need fitting or modelling for interpretation.
Our calculations reproduce the measured closing of the band gap at the bubble perimeter, without the necessity of fitting phenomenological model parameters.
Thus, we provide a robust \emph{ab initio} method to describe ZrTe$_5$, which is validated by STM     measurements.
Furthermore, the calculated equilibrium lattice parameters are also in agreement with values measured by x-ray diffraction of our sample.
This allows us to establish the electronic ground state of the system and map the topological phase diagram of ZrTe$_5$, revealing a STI phase at equilibrium.
Extended data of our calculations and an interactive way to browse them are available online at \href{https://tajkov.ek-cer.hu/zrte5phasediagram/}{tajkov.ek-cer.hu/zrte5phasediagram/} .

\section*{Experimental results}

\begin{figure}[h!]
\begin{center}
	\includegraphics[width = 0.99 \textwidth]{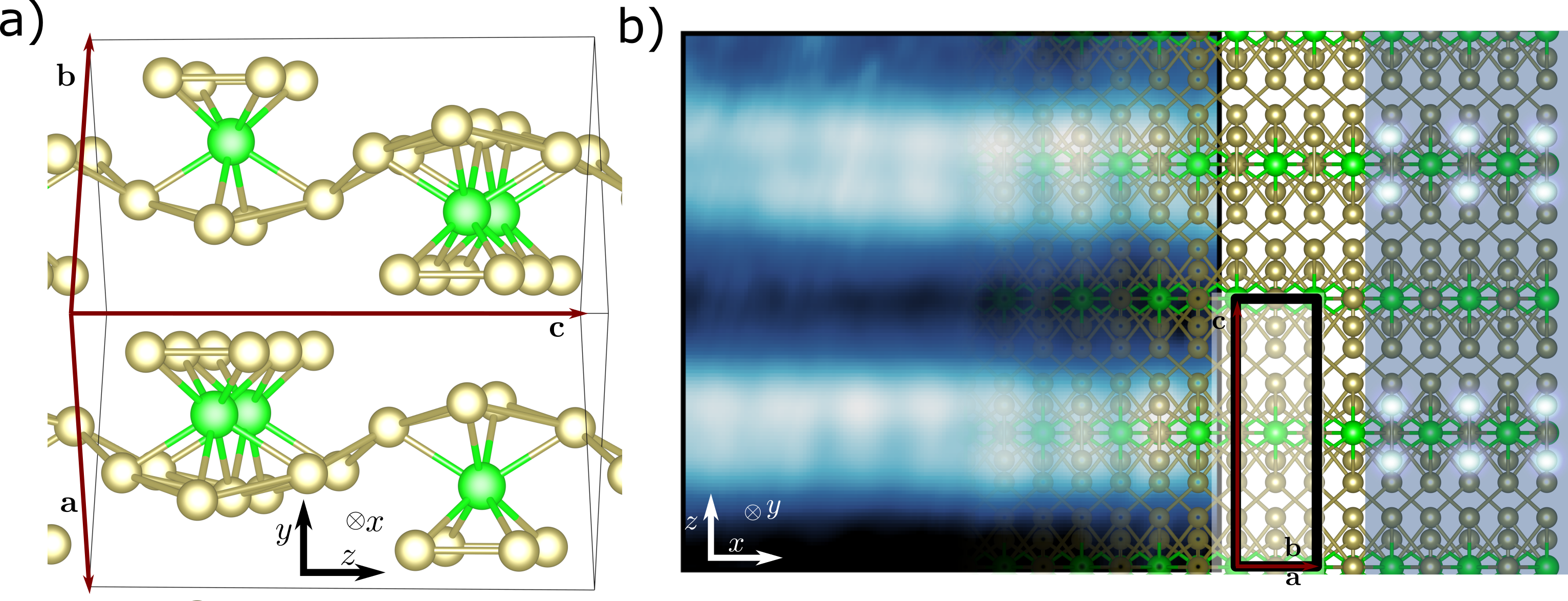}
	\caption{\textbf{Geometry of ZrTe$_5$.}
		\textbf{a)} The considered unit cell of the crystal.
		The wavelike Te-Te chains run parallel to the crystallographic $c$ direction and to the $z$-axis in the corresponding Cartesian coordinate system.
		\textbf{b)} Atomic resolution STM image of ZrTe$_5$ surface.
		The crystal structure, from the $(110)$ point of view, in the $x-z$ plane is superimposed in a fading manner.
		The STM image calculated by DFT is overlaid on the crystal structure (right side).
		It highlights the Te atoms on the sample surface which dominate the atomic resolution STM topography.
		STM image measured at 800 pA tunneling current, 400 mV sample bias, temperature 9 K.
		}
\label{fig:geom}
\end{center}
\end{figure}

The bulk zirconium pentatelluride (ZrTe$_5$) crystallizes in the layered orthorhombic crystal structure with space group \textit{Cmcm} (D$^{17}_{2h}$). Crossed trigonal prismatic ZrTe$_3$ chains run along the crystallographic $a$ direction and they are linked along the $c$-axis via parallel zig-zag chains of Te atoms, as it can be seen on Fig. \ref{fig:geom}, panel a).
The chains form two-dimensional sheets, stacked along the $y$-axis, forming a layered structure in the $x-z$ plane. The corresponding unit cell consists of 2 Zr atoms and 10 Te atoms. This is presented on Fig. \ref{fig:geom} panel b), where the atomic resolution STM image is matched with the computer generated image of the crystal in the $x-z$ plane. Each ZrTe$_5$ layer is nominally charge neutral, and the coupling between the layers is of van der Waals type \cite{weng2014transition}. 

The initial geometry was fully relaxed by \textit{ab initio} calculations (see section Methods for more details).
The starting point of the relaxation was the lattice constants provided by our X-ray diffraction measurements of the bulk sample.
The experimental and relaxed theoretical geometry parameters can be found in Table \ref{table:1}.


\begin{table}[h!]
\begin{center}
\scalebox{0.82}{%
\begin{tabular}{ l c c c c c c c c c} 
\hline
  & $a$ ($\AA$) & $b$ ($\AA$) & $c$ ($\AA$) & $y_\mathrm{Zr}$ & $y_{\mathrm{Te}_1}$ & $y_{\mathrm{Te}_2}$ & $z_{\mathrm{Te}_2}$ & $y_{\mathrm{Te}_3}$ & $z_{\mathrm{Te}_3}$ \\
\hline\hline
 Experimental & 1.994 (0.002) & 7.265 (0.005) & 13.724 (0.005)  & 0.3135 & 0.6725 & 0.9196 & 0.1497 & 0.2138 & 0.4341 \\
 DFT & 2.002 & 7.204 & 13.876  & 0.3136 & 0.6567 & 0.9293 & 0.1472 & 0.2059 & 0.4343 \\
\end{tabular}
}
\end{center}
\caption{Unit cell dimensions and positional parameters in fractional for ZrTe$_5$ as derived from the analysis of X-ray diffraction data at 300 K (calculated standard deviation in parentheses) and from relaxed DFT calculations.}
\label{table:1}
\end{table}


\begin{figure}[h!]
\begin{center}
	\includegraphics[width = 0.99 \textwidth]{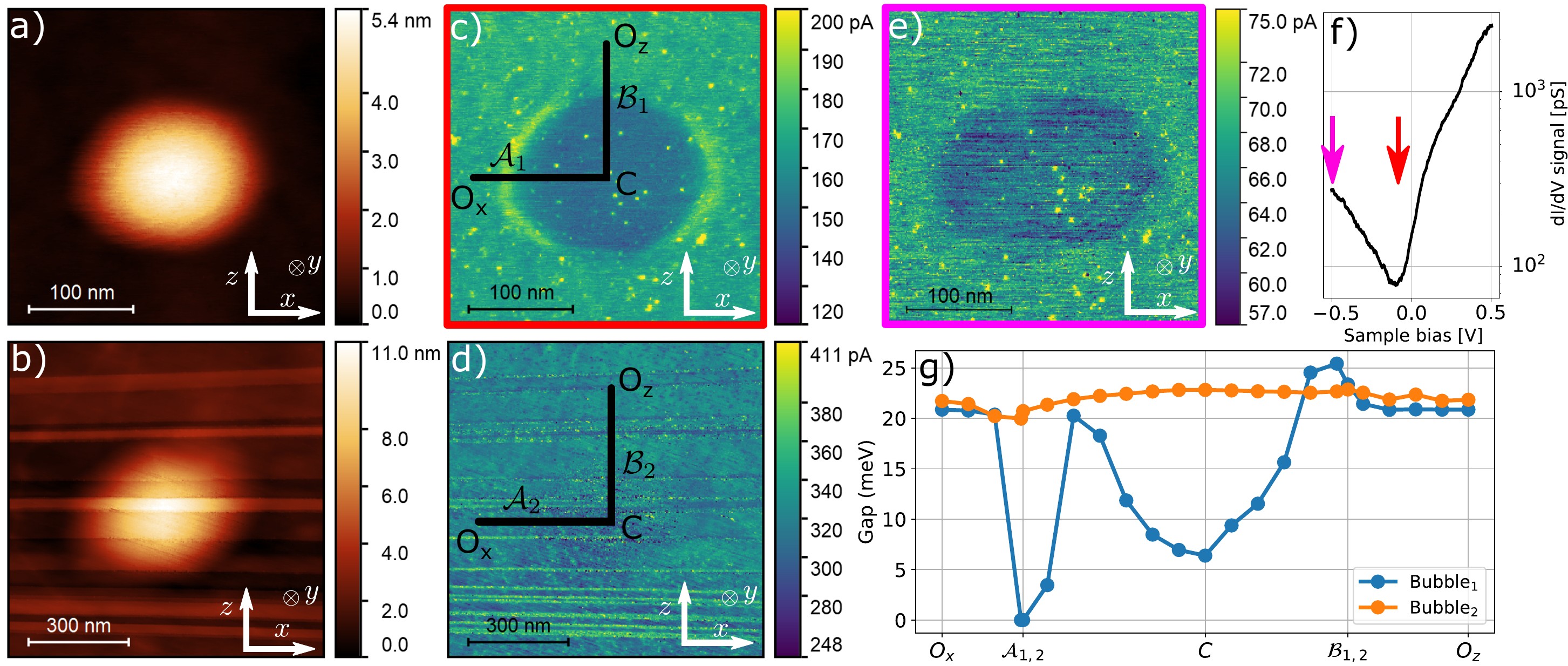}
	\caption{\textbf{Strain induced topological phase transition in ZrTe$_5$ bubbles.}
		\textbf{a)} STM topography image of Bubble$_1$ and \textbf{b)} of Bubble$_2$.
		\textbf{c)} Map of the $\mathrm{d}I/\mathrm{d}V$ signal in the same area as in a), measured within the gap.
		Bright dots on the $\mathrm{d}I/\mathrm{d}V$ image are point defects.
		\textbf{d)} Map of the $\mathrm{d}I/\mathrm{d}V$ signal in the same area as in b) of Bubble$_2$.
		\textbf{e)} Map of the $\mathrm{d}I/\mathrm{d}V$ signal in the same area as in panel c) but at -500 mV sample bias, deep in the valence band.
		\textbf{f)} $\mathrm{d}I/\mathrm{d}V$ spectrum measured on the sample surface far away from Bubble$_1$.
		Colored arrows highlight the sample bias used to measure the $\mathrm{d}I/\mathrm{d}V$ image in panel c) and e).
		The minimum of the $\mathrm{d}I/\mathrm{d}V$ signal corresponds to the gap and has a value of 78 pS.
		\textbf{g)} Band gap of ZrTe$_5$ resulting from \textit{ab initio} calculations, along the path depicted in c) and d).
		The gap, and as a consequence the LDOS, is modulated by the locally varying strain within the bubbles (see Fig.~\ref{fig:COMSOL}).
		STM measurement parameters: 300 K, 500 pA, sample bias for a)-d) -100 mV and -500 mV for panel e).
		}
\label{fig:STM}
\end{center}
\end{figure} 

Because ZrTe$_5$ decomposes under ambient conditions, we exfoliate the bulk crystals onto a Au(111) surface inside an inert glovebox environment, using the scotch tape method.
The samples are transferred by a vacuum suitcase to the UHV chamber of our STM, allowing the sample surface to remain pristine, as shown by the atomic resolution image in Fig.~\ref{fig:geom}b.
Bubbles form during exfoliation of van der Waals materials onto substrates and are predicted to contain inert hydrocarbon contamination~\cite{Haigh2012-ex}.
The a) and b) panels of Fig. \ref{fig:STM} show the STM topography image of two such bubbles.
They have an ellipsoid form and their geometry can be parameterized by the two semi-axis ($R_1$ and $R_2$) as well as the height ($h$) (see Table~\ref{table:2}).
Next to the topography images in panel c) and d) we present the corresponding measurements of the $\mathrm{d}I/\mathrm{d}V$ signal within the band gap, which is proportional to the local density of states (LDOS).
The middle of the gap is identified as the minimum of the $\mathrm{d}I/\mathrm{d}V$ signal, as shown in Fig.~\ref{fig:STM} f) (for more details see supplementary section \textit{S3.1}).
Focusing on Bubble$_1$ in panel c) a halo can be observed at the perimeter of the bubble, which is more intense perpendicular to the $x$ direction (point $\mathcal{A}_1$) and absent across the $z$ direction ($\mathcal{B}_1$).
This indicates an area with increased density of states within the gap, relative to the unstrained areas outside the bubble.
The anisotropy of the ring is a consequence of the highly anisotropic nature of the material, as we show below.
The increase in the density of states is only observed at energies within the gap.
This is illustrated by the fact that no halo is observed when mapping the $\mathrm{d}I/\mathrm{d}V$ signal at -0.5 V, away from the band gap well within the valence band (see Fig.~\ref{fig:STM} e)).
For Bubble$_2$ hosted by a thicker flake, shown in panels b) and d), the $\mathrm{d}I/\mathrm{d}V$ map displays no gap closing (point $\mathcal{A}_2$ in Fig.~\ref{fig:STM} g)), because the deformation values are much smaller in the thicker flake.

It is worth noting that the reduced LDOS within the bubble, stems from the lack of direct contact with the substrate.
The density of states of Au is orders of magnitude larger than that of the semimetallic ZrTe$_5$ and increases the LDOS measured in areas where the two materials are in close contact~\cite{Pan2018-jt,Feenstra2021-bf}.
This increase in LDOS is also present when measuring within the valence band (Fig.~\ref{fig:STM}e) but for thicker flakes it becomes much reduced (Fig.~\ref{fig:STM}d).

We calculated the strain pattern of the bubbles surfaces, using a unique technique that combines finite element calculations and density functional theory.
After we obtain the deformation of the bubbles, we can calculate the local electronic structure and the size of the gap along the path that is depicted in Fig.~\ref{fig:STM} panel c) and d) of the distorted crystal, using density functional theory.
As it can be seen in panel Fig.~\ref{fig:STM} g) DFT predicts a gap closure at the edge of Bubble$_1$, but only at the $\mathcal{A}_1$ point and not at $\mathcal{B}_1$.
The absence of the band gap means that the density of states must be higher in the area around the point, in agreement with the measurements.
Tracing the same path along the surface of Bubble$_2$, no gap closure can be seen, in good agreement with the measurement. 

\begin{table}[h!]
\centering
\begin{tabular}{ l | c c c c } 
  & $d [\AA]$ & $R_1 [\AA]$ & $R_2 [\AA]$ & $h [\AA]$ \\ \hline\hline
 Bubble$_1$ & 55 & 1015 & 813 & 49 \\ \hline
 Bubble$_2$ & 178 & 6000 & 4560 & 90 \\ \hline
\end{tabular}
\caption{Geometric parameters of Bubble$_{1, 2}$. Parameter $d$ denotes the thickness of the few layer ZrTe$_5$, $R_1$ and $R_2$ are the major and minor semi-axis and $h$ is the height of the bubble. All parameters are in the units of \AA.}
\label{table:2}
\end{table}

\section*{Numerical results}

\begin{figure}[t!]
\begin{center}
	\includegraphics[width = 0.99 \textwidth]{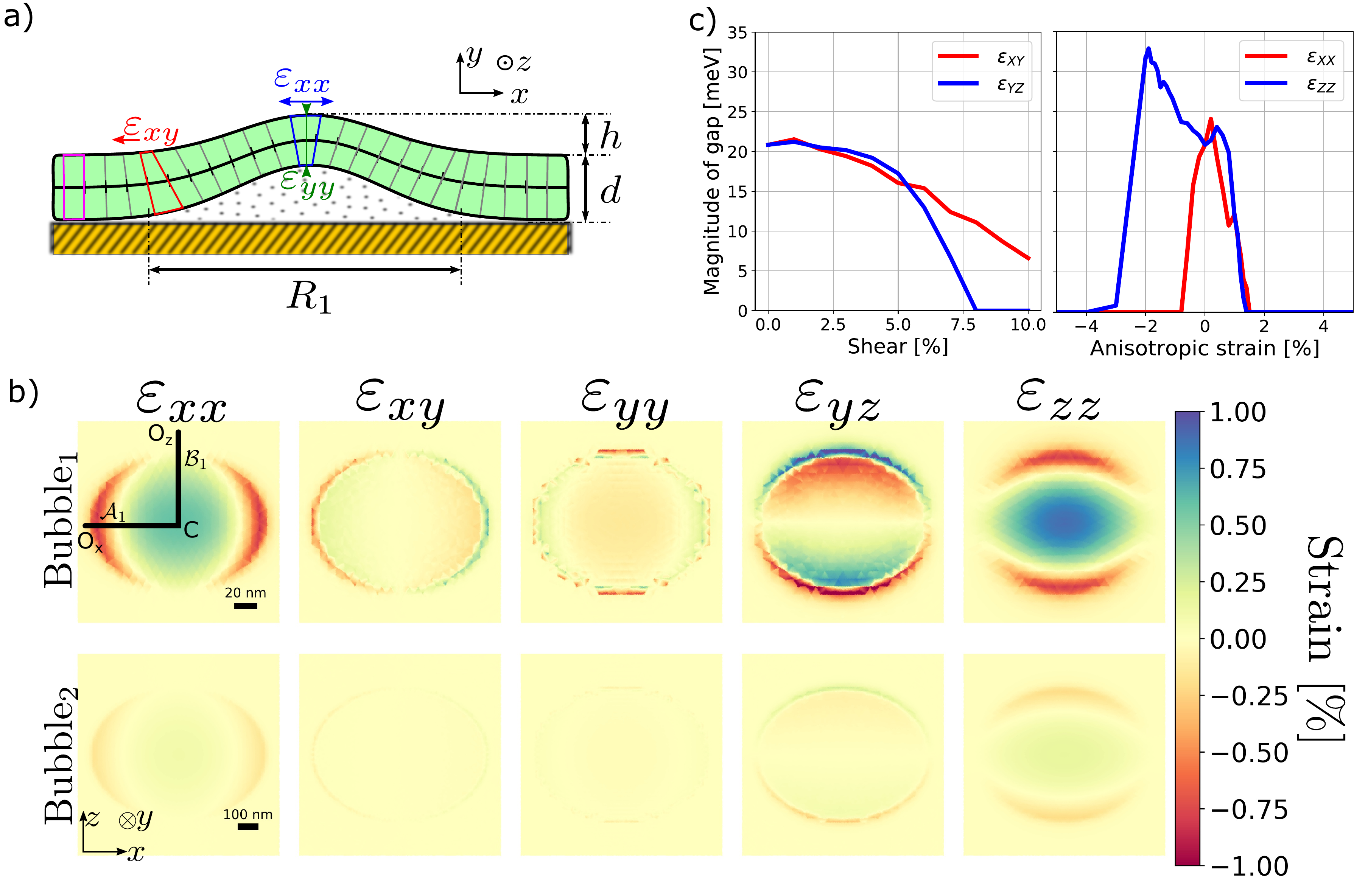}
	\caption{\textbf{Numerical results}
		\textbf{a)} A simplified view of the strain patterns in a bubble.
		The yellow, black dashed region indicates the gold substrate. The middle black solid line depicts the neutral plane, which shares the same geometrical parameters as the unstrained crystal. The different trapezoid quadrilaterals indicate the different strain patterns. We have also indicated the geometric parameters.
		\textbf{b)} Strain pattern on the top surface of the bubbles obtained by finite element calculations (COMSOL).
		This is the surface measured by STM.
		The top row corresponds to Bubble$_1$, the bottom row to Bubble$_2$.
		The different panels show different strain tensor components.
		Negative values correspond to compression.
		\textbf{c)} The influence of different strain components on the electronic properties. The magnitude of the gap as the function of the strain components, calculated by DFT.
		}
\label{fig:COMSOL}
\end{center}
\end{figure}

In order to avoid computationally intensive simulations for the bubbles containing a large number of atoms, we combined density functional theory and finite element calculations. In this method, the mechanical stress field of the bubbles was determined using the finite element method, where the bubbles being considered as continuous elastic materials. The obtained stress values were then used to perform DFT electronic structure calculations in bulk samples, with mechanical distortions corresponding to the local stress values. To apply this methodology, first, we must establish a proper description of the elasticity tensor of the material. For this we calculated the stiffness tensor by fitting the free energy of the distorted crystal via DFT.
For more details see Methods section.
The corresponding tensor elements are depicted in Table \ref{table:3} in Voigt notation\cite{mineralogical_society_1912}.
This stiffness tensor can then be used to describe the strain pattern of the bubbles, using finite element calculations.
We simulated the bubbles as a continuous anisotropic material, which can be characterized by the calculated stiffness tensor.
We matched the thickness of the sample to the experiments and applied hydrostatic pressure~\cite{khestanova2016universal} to the bottom of the system until we reached the measured height and reproduce the experimentally measured height profile.
For more information about the procedure see Methods and Section \textit{S2} of the supplement.
The calculated strain patterns can be seen in Fig.~\ref{fig:COMSOL}.
In panel a) we show a schematic representation of a bubble from a side view and denote the most important geometrical and strain parameters.
The yellow, diagonally dashed region denotes the substrate.
The solid black line in the middle of the bubble shows the neutral plain.
The different trapezoid quadrilaterals indicate the different strain patterns.
In Fig.~\ref{fig:COMSOL} panel b) we show the strain pattern for the two bubbles, plotting all relevant strain tensor elements on the surface of the bubble.
We have only omitted the presentation of the $\varepsilon_{xz}$ component as it is practically zero everywhere in the sample.
We show the exact numerical results in the supplementary material.
The first row corresponds to Bubble$_1$.
It can be observed that the strain patterns in the two bubbles show the same qualitative tendency, but the magnitudes are almost 5 times larger in the first bubble, due to the larger aspect ratio~\cite{khestanova2016universal}.

\begin{table}[!ht]
    \centering
    \begin{tabular}{|l|l|l|l|l|l|l|l|l|}
    \hline
    $C_{11}$ & $C_{22}$ & $C_{33}$ & $C_{12}$ & $C_{13}$ & $C_{23}$ & $C_{44}$ & $C_{55}$ & $C_{66}$ \\ \hline
        87.443  & 43.156  & 79.928  & 9.899  & 25.027  & 11.35  & 2.82  & 30.182  & 11.593 \\ \hline
    \end{tabular}
    \caption{Stiffness tensor obtained by DFT in Voigt notation and in the units of GPa.}
    \label{table:3}
\end{table}

As a final step using the strain field we obtained from the finite element method results we can calculate the band structure of the system under the influence of the strain patterns by using DFT.
In panel c) we present the effect of four different strain components on the bandgap of the crystal.
In the left side of subfigure c) we show the elements that are responsible for the in-plane shear.
It is clear that both decrease the band gap, but the $\varepsilon_{xy}$ component has a smaller contribution compared to $\varepsilon_{yz}$.
Furthermore, it can be clearly observed that the magnitude required to close the gap is as high as 7.5 $\%$, which is too large to explain the effect observed on Bubble$_1$.
We can find the explanation to the LDOS halo observed in the meauseremnts by looking at the curves in the adjacent panel.
The solid blue line denotes the effect of the $\varepsilon_{zz}$ component on the band gap.
Positive strain values close the gap at around 1.5 $\%$, but the negative values first widen the gap and after 2$\%$ start closing it, eventually closing at around 4$\%$.
We found that $\varepsilon_{xx}$ and $\varepsilon_{zz}$ behave identically for positive values, only showing a drastic difference for compression, namely that $\varepsilon_{zz}$ closes the gap for much reduced strain values.

This allows us to explain the increase in LDOS at the Bubble$_1$ perimeter as a closing of the gap.
We consider the calculated gap values along the path in Fig. \ref{fig:STM} panel g).
Starting from the equilibrium $O_x$ point along the $x$-axis and moving towards the $\mathcal{A}_1$ point, the $\varepsilon_{xx}$ strain element becomes a large enough negative number to close the gap. This magnitude then goes to zero, elevates again, becomes a relatively smaller positive value, but never reaches the magnitude that closes the gap again.
After we reach the center of the bubble (point $C$) and turn upwards along the $z$-axis the $\varepsilon_{zz}$ becomes smaller, reaches zero, and then becomes a large negative value.
But as $\varepsilon_{zz}$ has a different role in the manipulation of the electronic structure of the crystal it slightly widens the gap in the $\mathcal{B}_1$ points compared to the $O_z$ points, where it reaches equilibrium again.
As for the second bubble, the explanation is much simpler, the magnitude of the strain components never reaches a sufficiently high value to have an observable influence on the gap.
In the supplementary material we present the exact values for every element of the strain tensor along the whole path (see SI S3.2).

\begin{figure}[ht!]
\begin{center}
	\includegraphics[width = 0.8 \textwidth]{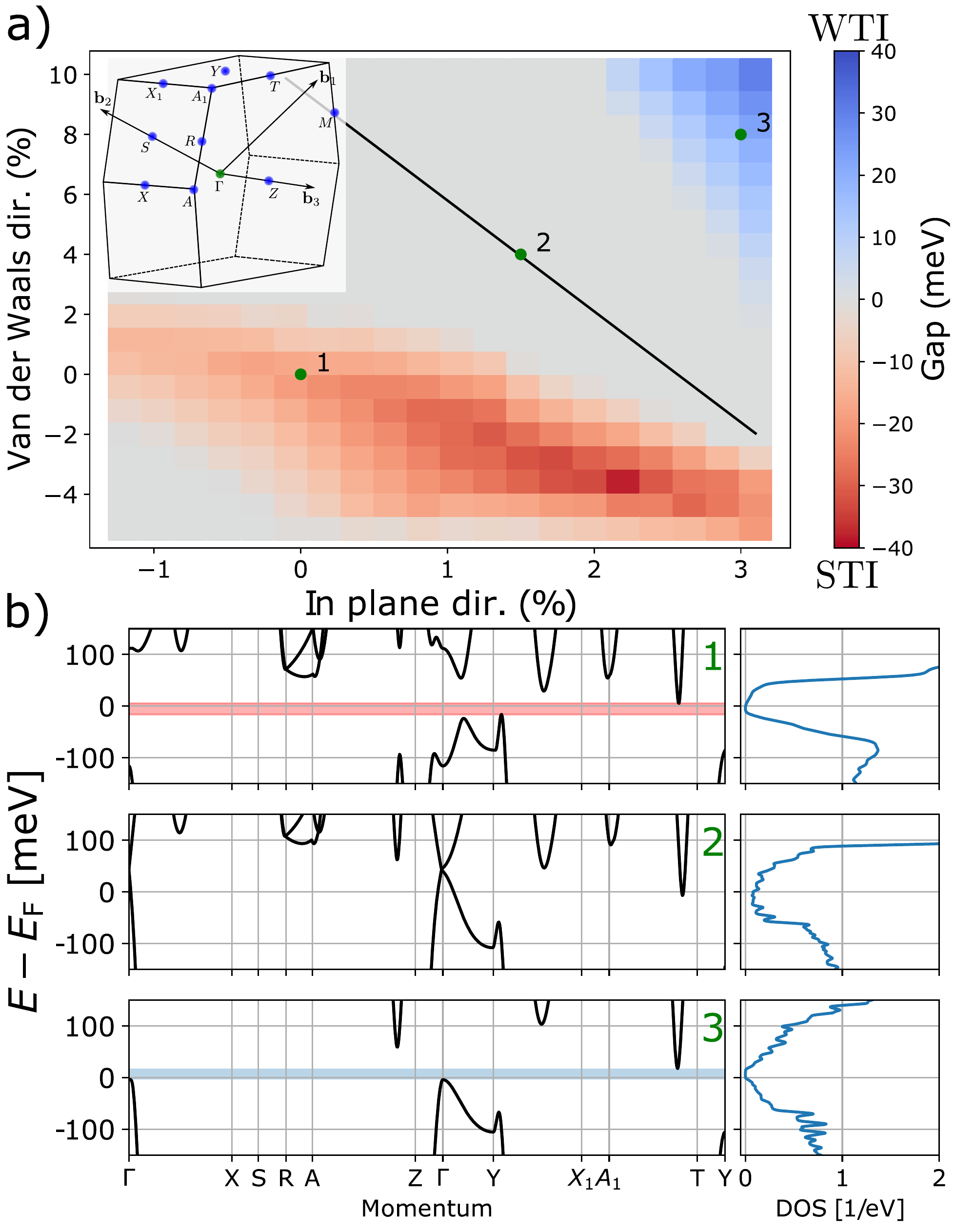}
	\caption{\textbf{Topological phase diagram of ZrTe$_5$.}
		\textbf{a)} The phase diagram of the electronic structure of the crystal under mechanical strain. 
		At every point the size of the gap was calculated and a sign has been assigned to it according to the topological favor of the gap.
		The negative gap corresponds to the strong topological insulating phase, while the positive gap to the weak topological phase.
		The black solid tentative line indicates the boundary where the Dirac cones in the $\Gamma$ point touch each other.
		The green dots assigned with a number denote the corresponding band structure in the b) subfigure. The inset shows the corresponding Brillouin zone indicating the high symmetry points.
		\textbf{b)} The calculated band structure along the path of the high symmetry points. The opaque band shows the size of the gap and color indicates the topological favor.
		}
\label{fig:Phase}
\end{center}
\end{figure}

Since our \textit{ab inito} calculation reproduces the effect of the complex strain pattern on the ZrTe$_5$ band gap, we can take one step further and map the phase diagram of the system.
The first panel in Fig. \ref{fig:Phase} shows the contour map of the band gap size ($E_\mathrm{g}$) under different mechanical strains.
The horizontal axis indicates the in-plane~\footnote{The lattice vectors $\textbf{a}$ and $\textbf{b}$ were distorted isotropically} strain from -1.2 $\%$ to 3.1$\%$ in 20 steps, while the vertical axis corresponds to the van der Waals direction ($y$-axis) from -5 $\%$ to 10 $\%$ in 20 steps. 
At every point a sign has been assigned to the gap as the $\mathbb{Z}_2$ invariants were calculated, positive (negative) sign indicates WTI (STI).
The phase diagram shows three main domains.
Around the equilibrium the system is a STI (1) and it can be tuned to the WTI (3) phase through a conducting phase (2).
The black line in the conducting phase shows where the Dirac cones in the $\Gamma$ point touch each other.

The conclusion that the equilibrium state is a STI, is supported by our $\mathrm{d}I/\mathrm{d}V$ measurements.
For a STI, the LDOS within the gap should not go down to zero because of the presence of the topological surface state~\cite{Shen2017-ct}.
Taking into account the noise level in our instrument, a value of zero LDOS would correspond to a tunneling conductance value lower than 1 pS.
Compared to this, the conductivity at 300 K inside the gap is 78 pS (see Fig.~\ref{fig:STM}f) and at 9 K is 74 pS (see supplementary section \textit{S3.1}).

Fig.~\ref{fig:Phase} b) shows three typical band structures at points 1, 2 and 3 marked in panel a), and total density of states in 1/eV units.
The first band structure shows a strong topological insulating phase, the red opaque band highlights the 22 meV band gap.
As we go towards point (2) we reach the grey area where the band gap is closed but there is no touching of the bands.
At point (2) the bands corresponding to the massive Dirac fermions touch each other around $\Gamma$.
As we go further towards point (3) the Dirac cones open up and the band between the high symmetry points $A_1$ and $T$ lift up from the Fermi level.
This opens a gap in the weak topological insulating phase.

We repeated our calculations using another commonly used DFT code, VASP, that resulted in a similar phase diagram to the one presented in Fig.~\ref{fig:Phase}. Besides the similarities, there are also differences that are instructive and bear information. A detailed discussion of these calculations and their implications can be found in the supplementary information section \textit{S4}.

\section*{Summary}


ZrTe$_5$ is a unique material, with electronic properties that are extremely sensitive to its lattice parameters.
This over-sensitivity has led to controversies in the crystal's literature in recent years.
On the other hand, the sensitivity makes it a perfect candidate for strain engineering the material across a topological phase transition.
In this contribution we have combined  density functional theory and finite element calculations with direct measurements of the charge density, allowing us to identify the mechanical deformation needed to induce topological phase transitions in the material.
We showed that shear in the plane perpendicular to the van der Waals direction could also lead to a phase transition.
Our results pave the way for a robust understanding of the band structure and topological properties of ZrTe$_5$, enabling future investigations into the physics of topological phase transitions as well as applying these in electronic devices.

\section*{Author contributions}

ZT, DN, OL and JK performed the \textit{ab initio} calculations, with assistance from PV.
ZT performed calculations of the mechanical properties (COMSOL), with assistance from PS.
KK was responsible for the sample preparation and performed the STM measurements.
ZEH measured and evaluated the the x-ray diffraction data.
LT contributed to data interpretation.
ZT and PNI wrote the manuscript, with input from all authors.
PNI conceived and coordinated the project.

\section*{Methods}

\textbf{DFT}

The optimized geometry and electronic properties of the crystal were obtained by the SIESTA implementation of density functional theory (DFT)\cite{artacho2008siesta,soler2002siesta,garcia2020siesta,fernandez2006site}. SIESTA employs norm-conserving pseudopotentials to account for the core electrons and linear combination of atomic orbitals to construct the valence states. The generalised gradient approximation of the exchange and the correlation functional was used with Perdew–Burke–Ernzerhof parametrisation\cite{perdew1996generalized} and the pseudopotentials optimised by Rivero \textit{et al}.\cite{rivero2015systematic} with a double-$\zeta$ polarised basis set and a realspace grid defined with an equivalent energy cutoff of 350 Ry for the relaxation phase and 900 Ry for the single-point calculations. The Brillouin zone integration was sampled by a 30$\times$30$\times$18 Monkhorst–Pack $k$-grid for both the relaxation and the single-point calculations.\cite{monkhorst1976special} The geometry optimisations were performed until the forces were smaller than 0.1 eV nm$^{-1}$ . The choice of pseudopotentials optimised by Rivero \textit{et al}. ensures that both the obtained geometrical structures and the electronic band properties are reliable. After the successful self consistent cycles the necessary information was obtained by the sisl tool \cite{zerothi_sisl}. The spin orbit coupling was taken into account in the single point calculations.

The simulated STM image was obtained for a three layers thick sample, that we cut from the original bulk crystal in the (110) orientation. We used Monkhorst Pack resolution $20 \times 12 \times 1$ in the geometry relaxation this case, where the samples were separated with a 40 Å thick vacuum in the perpendicular direction. The simulated STM image was obtained by the tools developed by the SIESTA developers, where the position of the tip was 2.5 Å away from the position of the topmost Te atom.


\textbf{Stiffnes tensor}

We obtained the stiffness tensor elements by fitting the free energy change of the crystal under mechanical deformations. The change in the free energy of the crystal is a quadratic function of the strain tensor \cite{landau_lifsic8}. The procedure presented here is the most recent way to precisely determine the elements of the stiffness tensor using DFT \cite{wang2021vaspkit}.

The general form of a deformed crystal is the following \cite{landau_lifsic8}:

\begin{equation}
    F=\frac{1}{2}\lambda_{klmn}u_{kl}u_{mn},
\end{equation}
where $\lambda_{klmn}$ is the elastic modulus tensor, $u_{ij}$ is the strain tensor. The general expression for the free energy in the orthorhombic system is
\begin{equation}
\label{eq:free_energy}
\begin{split}
F & = \frac{1}{2}\lambda_{xxxx}u_{xx}^2+\frac{1}{2}\lambda_{yyyy}u_{yy}^2+\frac{1}{2}\lambda_{zzzz}u_{zz}^2 + \lambda_{xxyy}u_{xx}u_{yy}+ \\
 + & \lambda_{xxzz}u_{xx}u_{zz}+\lambda_{yyzz}u_{yy}u_{zz}+2\lambda_{xyxy}u_{xy}^2+2\lambda_{xzxz}u_{xz}^2+2\lambda_{zyzy}u_{zy}^2.
\end{split}
\end{equation}
It contains nine moduli of elasticity \cite{landau_lifsic8}. By applying mechanical strain to the relaxed geometry in the \textit{ab initio} calculations using the free energy obtained by SIESTA software we can fit Eq. (\ref{eq:free_energy}) to get the different moduli.

\textbf{Finite element method}

The corresponding strain patterns were calculated by numerically solving the three-dimensional equation of motion by the finite element method (FEM). The FEM calculation was performed by the MEMS Module of the COMSOL Multiphysics\textsuperscript{\textcopyright} \cite{multiphysics1998introduction} 5.6 software package. The bubbles were simulated by a 20000$\times$20000 $\AA^2$ block. The thickness of the block was matched to the sample size in the STM measurements. The sides and the bottom of the block were fixed and we applied hydrostatic pressure on the bottom of the block on an ellipse shaped part of the bottom to match the shape of the bubbles in the STM measurements. The pressure was chosen to match the height of the bubbles.

\textbf{Sample preparation and STM measurements}

The ZrTe$_5$ crystals were purchased from hqgraphene.com and exfoliated onto gold substrates~\cite{Magda2015-nn}, inside an inert glovebox environment.
The samples were transferred to the chamber of the UHV STM, via a vacuum shuttle.
Atomic resolution images show that the ZrTe$_5$ crystal surface remains pristine after transfer into the STM chamber (see Fig. \ref{fig:geom}).

STM measurements were performed using an RHK PanScan Freedom microscope at 300 K and 9 K temperatures in UHV, at a base pressure of 5$\times$10$^{-11}$ Torr.
STM tips were prepared by mechanically cutting Pt/Ir (90\%/10\%) wire.
We have used a large working distance optical microscope to place the STM in the vicinity of selected ZrTe$_5$ crystals.
$\mathrm{d}I/\mathrm{d}V$ spectra were measured using a Lock-In amplifier, with a reference frequency of 1.372 kHz and a bias modulation of 5 mV at 9 K and 30 mV at 300 K.

\textbf{X-ray diffraction}

X-ray diffractometry measurements were performed in parallel geometry using a Bruker AXS D8 Discover diffractometer equipped with Göbel-mirror and a scintillation detector with Cu K$\alpha$ radiation.
The X-ray beam dimensions were 1 mm $\times$ 5 mm, the 2 step size was 0.02$^{\circ}$, scan speed 6 sec/step.
We used the Diffrac.EVA program and the ICDD PDF database for phase identification.

\section*{Acknowledgements}
	
	The work was conducted within the framework of the Topology in Nanomaterials Lendulet project, Grant No. LP2017-9/2017, with support from the European H2020 GrapheneCore3 Project No. 881603.
	Financial support from \'{E}lvonal Grant KKP 138144, NKFIH OTKA grant K132869 and TKP20121 NKPA grant is also acknowledged.
	PS, acknowledges KIF\"U for awarding access to computing resources.
	PV and LO acknowledge the support of the Janos Bolyai Research Scholarship the Bolyai+ Scholarship of the Hungarian Academy of Sciences. LO acknowledges financial support from NKFIH OTKA grant FK124723 and K131938.
	LO and JK acknowledges the support from the Ministry of Innovation and Technology for the Quantum Information National Laboratory.
    
\newpage

\bibliography{zrte5_cikkek}

\end{document}